\documentstyle[12pt]{article}

\begin{document}

\begin{center}
{\Large{
 {\bf{                 RELAXED STATES OF \\ A MAGNETIZED  PLASMA  \\
                       WITH MINIMUM DISSIPATION\\
}}}}
\end{center}

\vskip 0.5cm

\begin{center}

{\bf{B. Dasgupta$^{*}$, P. Dasgupta$^{**}$,\\
M. S. Janaki$^{*}$, T. Watanabe$^{***}$ and T. Sato$^{***}$}}\\
$*$Saha Institute of Nuclear Physics\\
I/AF, Bidhannagar, Calcutta 700 064, India\\
$**$Department of Physics, University of Kalyani, Kalyani  741235, India\\
$***$ Theory and Computer Simulation Center\\
National Institute for Fusion Science, Toki, Gifu 509-52, Japan

\end{center}

\vspace{10ex}
\begin{center}
{\bf{Abstract.}}
\end{center}
\vspace{5ex}

Relaxed state of a slightly resistive and turbulent
magnetized plasma is obtained by invoking the Principle of Minimum
Dissipation which leads to 
$$ \nabla \times \nabla \times \nabla \times {\bf B} = \Lambda{\bf{ B}}. $$

\noindent
A  solution of the above equation
is accomplished using the analytic continuation of the Chandrasekhar-Kendall
eigenfunctions in the complex domain. 
 The new features of this theory is to show (i)
a single fluid can relax to an MHD equilibrium
which can support pressure gradient
even without a long-term coupling between  mechanical flow and  magnetic field
(ii) field reversal (RFP) in states that are not force-free.

\newpage
In the well-known theory of relaxation of magnetoplasma, 
Taylor\cite{Tay}
proposed that the process of relaxation is governed
by the principle of minimum total magnetic energy and
invariance of total (global) magnetic helicity  
$ K = \int_V {\bf{A}}\cdot {\bf{B}}dV$
\noindent where the integration is over the
entire volume, the latter being the most significant 
invariant in the theory of 
relaxation. Accordingly the relaxed state of a 
magnetoplasma satisfies the corresponding Euler-Lagrange 
equation 

\begin{equation} \nabla \times {\bf{B}} = \lambda {\bf{B}} 
\end{equation}

\noindent with constant $\lambda$, and, consequently,  
is a force-free state. Taylor's theory is quite successful 
in 
explaining a number of experimental results, including 
those of RFP. 
However, relaxed states as envisaged by Taylor, have only 
zero
pressure gradient.

Extensive numerical works by Sato and his collaborators 
have
established \cite{Sato1},\cite{Sato3} the existence of 
self-organized states with
finite pressure, i.e. these states are governed by the
magnetohydrodynamic force balance relation, namely,
${\bf{j}}\times{\bf{B}} = \nabla p$, rather than   
 ${\bf{j}}\times{\bf{B}} = 0$. 
Recently, it has been 
demonstrated both by numerical simulation\cite{Sato2}
and by experiments\cite{Ono}
that counter-helicity merging of two spheromaks can
produce a Field-Reversed Configuration (FRC). The FRC 
has zero toroidal magnetic field and the plasma 
is confined entirely  by poloidal magnetic field. It has 
a finite pressure with a relatively
high  value of $\beta$.  It may be concluded that
FRC, with its non zero perpendicular component of current, 
is a relaxed state and it is a distinctly non-force free 
state. 
From the point of view of plasma relaxation, the formation
of FRC through the counter-helicity merging of two
spheromaks is a unique process where a non-force free state 
emerges 
from the fusion of two Taylor states. The conclusion is 
that there exists a general class of 
relaxed states  which are not always 
force-free, and Taylor's force-free states constitute only
a subclass of this wider class.
While Taylor states do
not support any pressure gradient, equilibrium obtained 
from the
principle of minimum energy accommodates pressure gradients 
only in
presence of flow.
Several attempts\cite{Ham},\cite{Avi} have been made in the  past
to obtain relaxed states which could support finite
pressure gradient, a large number of them  
making use of the coupling of the flow with magnetic 
field\cite{Finn}-\cite{Stein}.

The principle of "minimum rate of entropy production",
formulated by Prigogine\cite{Prig} and others, is
believed to play a major role in many
problems of irreversible thermodynamics. 
Dissipation, along with nonlinearity, is ubiquitous 
in systems which evolve towards  self-organized
states. Another closely
 related concept, the principle of minimum dissipation rate
was used for the first time by Montgomery and 
Phillips\cite{Mont} 
in an MHD problem to understand the steady state profiles 
of RFP configuration under the constraint of constant rate 
of supply 
and dissipation of helicity 
the usual physical boundary conditions for a conducting 
wall.
It may be pointed out that the principle of minimum 
dissipation 
was also discussed by
Chandrasekhar and Woltzer\cite{CK2} in a sequel to
the complete general solution of the force-free equation by 
Chandrasekhar and Kendall \cite{CK1}.
The minimum dissipation rate hypothesis was later used by a 
number of 
authors \cite{Wang},\cite{Bevir} to predict the current and 
magnetic field 
profiles of driven dissipative systems.

This paper deals with the question of determining the field
configurations assumed by a magnetofluid in a relaxed state  
{\it{in absence of any external
fields}}, while maintaining that the relaxation is governed 
by 
the hypothesis of minimum rate of energy dissipation.  It 
is our conjecture
that relaxed states could be characterized as the states of 
minimum
dissipation rather than states of minimum energy.
  The novel feature of our work
is to show that it is possible for a {\em{single}} fluid to 
relax to
an MHD equilibrium with
a magnetic field configuration which can support
pressure gradient, even without a long-term coupling 
between the 
flow and the magnetic field. In a recent work, 
Steinhauer\cite{Stein}
has claimed that single fluid MHD theory can admit only a 
force free state and
one need  to take recourse to a two fluid theory so that 
electromechanical
coupling produces pressure gradient and a non force free 
state.
Our work establishes  that none of these requirements
 need be satisfied to obtain a relaxed state of the desired 
kind. 

In what follows we derive the Euler-Lagrange equation from 
a variational
principle
with minimum energy dissipation and conservation of total 
magnetic 
helicity, solve the equation in terms of the analytically 
continued 
Chandrasekhar-Kendall eigenfunctions, discuss 
the important role played by the boundary conditions, 
and present our results for the flux, field
reversal parameter and pinch parameter. We also compute
the helicity integral, and show the plots of magnetic 
field, current, 
and pressure profiles. The field reversal parameter from 
our theory
 is definitely in better agreement with the experimental 
value than 
 what is obtained from Taylor's theory.

We consider~~a closed system of  an incompressible, 
resistive~~ 
magnetofluid, without any
mean flow velocity, described by the standard MHD equations 
in presence
of a small but finite resistivity $\eta$.  
In the absence of any externally imposed electric fields, 
the ohmic dissipation rate $R $ is itself
a time varying quantity.  However, it is possible to find 
constraints
that are better preserved than the rate of energy 
dissipation, so that 
the system self-organizes to certain relaxed states, which 
remain stable 
 over time scales short compared to ohmic dissipation time. 
In this case,
 helicity still serves to hold as a good
constraint as it decays at a time scale much slower 
in comparison to the decay time scale of the rate of energy 
dissipation
as is evident from the simulation works of Zhu et. al. 
\cite{Sato3}.  
In the following,
 we compare the decay rates of the energy dissipation rate 
and helicity.
The former is obtained as
$$\frac{dR}{dt} = 2\frac{\eta^2}{S^2}\sum_k k^4 {\bf b}_k^4 
$$
while the latter turns out to be
$$\frac{dK}{dt}=-2\frac{\eta}{S}\sum_k k{\bf b}_k^2.$$ 
From the above two equations, we see that the decay rate of 
energy dissipation
is once again O(1) at scale lengths for which $k \approx 
S^{\frac{1}{2}} $.
But at these scale lengths, helicity dissipation is only 
O$(S^{-1/2}) <<1 $.
Thus, we may expect that in presence of
small scale turbulence, the rate of energy dissipation 
decays at a 
faster rate than helicity.

We therefore minimize the ohmic dissipation
$R = \int \eta {\bf{j}}^2 dV $  subject to the
constraints of helicity $ \int {\bf A} \cdot {\bf B} dV $.
The variational equation is given by

\begin{equation} \delta \int \left(\eta{\bf{j}}^2 + 
{\overline \lambda}{\bf{A}}\cdot
{\bf{B}}\right)dV = 0 \end{equation}

\noindent where $\overline \lambda$ is Lagrange's  
undetermined multiplier.
The variation can be shown to lead to the Euler-Lagrange 
equation 
\begin{equation} \nabla \times \nabla \times \nabla \times 
{\bf B}  = {\Lambda}{\bf{B}}
 \end{equation}

\noindent where, $\Lambda ={\overline\lambda} /\eta $ is a 
constant.
The surface terms in the equation vanish if we consider the 
boundary condition
$\delta {\bf A} \times {\bf n} = $ as well as ${\bf 
j}\times {\bf n} = 0 $, which
is the physical boundary condition we will impose in the 
problem.

We like to emphasize that eq.(3) is a general equation
which embraces the Woltzer-Taylor equation ( i.e. eq. (1) ) as a special 
case.
Now we proceed to construct a solution of eq.(3)
and show that the general solutions can have
 ${\bf{j}}\times{\bf{B}}\neq 0$. In other
words, eq.(3) may lead to a non force-free state.

The solution of eq.(3) can be constructed
using the Chandrasekhar-Kendall (CK) eigenfunctions. 
Chandrasekhar and Kendall's solution\cite{CK1} of the 
equation 
 $\nabla\times{\bf{B}} = \lambda~{\bf{B}}$
 can be written (with three parameters $\mu, m, k$ in
cylindrical coordinates), as,

\begin{equation} {\bf{B}}(\mu, m, k)
         =  \lambda\nabla\Phi\times\nabla z + 
             \nabla\times(\nabla\Phi\times\nabla z) 
\end{equation}

\noindent where, $\Phi = J_m (\mu r)~exp[i(m\theta - kz)]$
with $\lambda^2 = \mu^2 + k^2$. Here,
$J_m$ is a Bessel Function of order $m$ and the value of 
$\mu$ in the argument
is determined from the boundary condition at $r = a$, which 
is 
given as $({\bf{\hat{n}\cdot {\bf B}}})_{r=a} = 0$

Analytic continuation of the above solution for complex
values of $\mu$ (or $k$) is straightforward.  
For real values of $\lambda$ (and hence of $\mu$ and $k$) 
the
operator $(\nabla\times)$ has been proved to be self-
adjoint,
but not so in the larger space spanned by the analytically
continued CK solutions.

We introduce the complex parameters
\begin{equation}
\mu_n = \left[(\mu^2 + k^2)exp(4n\pi~i/3) - 
k^2\right]^{1/2} , n=1,2
\end{equation}
\noindent so that $ \mu_n^2 + k^2 =\lambda^2 \omega^{2n} $, 
$\omega = exp (2\pi i/3) $, and define
\begin{eqnarray}
{\bf{B_1}} &=& {\bf{B}}(\mu,m,k)=
\lambda \nabla\Phi\times\nabla z + 
             \nabla\times(\nabla\Phi\times\nabla z) 
\nonumber \\
{\bf{B_2}} &=& {\bf{B}}(\mu_1, m, k)=
             \lambda \omega\nabla\Phi_1\times\nabla z + 
             \nabla\times(\nabla\Phi_1\times\nabla z) 
\nonumber \\
{\bf{B_3}} &=& {\bf{B}}(\mu_2, m, k)=
             \lambda \omega^2\nabla\Phi_2\times\nabla z + 
             \nabla\times(\nabla\Phi_2\times\nabla z)
\end{eqnarray}

\noindent In the last two expressions above, $\Phi_1$ and 
$\Phi_2$
are obtained from $\Phi$ by replacing $\mu$ by $\mu_1$ and 
$\mu_2$
respectively. 

A solution of  eq.(3) can now be obtained as a linear 
combination
of ${\bf{B_1}},~ {\bf{B_2}},~ {\bf{B_3}}$ :

\begin{equation} {\bf{B}} = \alpha_1 {\bf{B_1}} + 
\alpha_2 {\bf{B_2}} + \alpha_3 {\bf{B_3}}
\end{equation}
\noindent where $\alpha_i$ are constants, with at least two
of them non-zero.
It can be easily demonstrated that the expression 
for {\bf{B}} given in (7)
is a solution of eq.(3) with $\Lambda = \lambda^3 $.

A reasonable boundary condition is to assume a perfectly 
conducting wall, so that

\begin{equation} 
{\bf{B}}\cdot{\bf{n}}=0,~~{\bf{j}}\times{\bf{n}}=0
~~~~ {\rm {at}}~~~ r = a \end{equation}

The boundary conditions given by eq. (8) suffice to fix the arbitrary
constants
\begin{eqnarray}
\frac{\alpha_2}{\alpha_1}&=&
 -\frac{\omega^2(B_{1\theta}B^*_{2z}-B^*_{2\theta}B_{1z})\mid_{r=a}}
{(B_{2\theta}B^*_{2z}-B^*_{2\theta}B_{2z})\mid_{r=a}}   \\
\alpha_3 &=&\alpha_2 ^*
\end{eqnarray}

 The magnetic fields at the boundary $r=a $ have to obey the following
relation  for non-trivial values of the constants $\alpha_i $
\begin{equation} 2B_{1r}Im(B_{2\theta}B^*_{2z})-2 
B_{1\theta}
Im(\omega^2 B_{2r}B^*_{2z})
+2 B_{1z}Im(\omega^2 B_{2r}B^*_{2\theta})=0 \end{equation}

\noindent From eq. (6), it is evident that $B_2$ and $B_3$ are complex conjugate of one
another. This, together with the relations obtained in eq. (10), shows 
that the magnetic field given by eq. (7) is a real field.
 We also list the following expressions for the 
$m=0,k=0$ state 
(cylindrically symmetric state) obtained from eqs. (4)-(7) 
:
\begin{eqnarray*}
B_r &=& 0 \\
B_{\theta} &=& \lambda^2\alpha_1\left[J_1(\lambda r)+2 Re 
\left(\frac{\alpha_2}{\alpha_1}
\omega^2 J_1(\lambda \omega r)\right) \right] \\
B_z &=& \lambda^2\alpha_1\left[J_0(\lambda r)+2 Re \left( 
\frac{\alpha_2}{\alpha_1}
\omega^2 J_0(\lambda \omega r)\right)\right] \\
\end{eqnarray*}

For a given value of $m$ and $ka$, the value of $\lambda a 
$ can be obtained
from the boundary condition given by eq. (11).  
It is to be noted that for the cylindrically symmetric 
state
the boundary condition is trivially satisfied and hence 
does not 
determine $\lambda a$.
It can be easily proved 
 that the state of minimum dissipation is equivalent to the 
state of minimum
value of $\Lambda $.
To get the numerical value of $\lambda$ for  $m \neq 0$, we 
solve numerically eq.(11) 
 and obtain $\lambda a = 3.11 $
and $k a = 1.23 $ as the minimum values of $\lambda a $ and 
$ka $
for the $m=1 $ state.\\
The only undetermined constant in eq. (7) is the value of 
$\alpha_1 $
(the value of the field amplitude) which can be determined 
by 
specifying the toroidal flux $\Phi_z $.  The $m=k=0 $ state 
is responsible
for non-zero values of toroidal flux which is obtained as
\begin{equation}
\Phi_z = 2 \pi \alpha_1 \lambda a \left[ J_1(\lambda a) + 2 
Re [\frac{\alpha_2}{\alpha_1}
\omega J_1(\lambda \omega a)] \right]  \end{equation}
\noindent A couple of dimensionless quantities that have 
proved useful
in describing laboratory experiments are the field reversal 
parameter $F ={B_z(a)}/{<B_z>} $ and the pinch parameter 
$\Theta ={B_{\theta}(a)}/{<B_z>}$,
 where $<..>$ represents a volume average.
After substituting the expressions for $B_z(a)$
etc, we get
\begin{eqnarray}
F &=& \frac{\lambda a}{2}\frac{J_0(\lambda a)+2 Re \left 
[(\alpha_2/\alpha_1)
\omega^2 J_0(\lambda \omega a) \right ] }
{J_1(\lambda a)+2 Re \left [(\alpha_2/\alpha_1)\omega 
J_1(\lambda \omega a)\right ]} \\
\Theta &=& \frac{\lambda a}{2}\frac{J_1(\lambda a)+2 Re 
\left [ (\alpha_2/\alpha_1)
\omega^2 J_1(\lambda \omega a) \right ] } 
{J_1(\lambda a)+2 Re \left [(\alpha_2/\alpha_1)\omega 
J_1(\lambda \omega a)\right ] }
\end{eqnarray}

The pinch ratio $\Theta $ is related to the ratio of the 
current and flux and is a physically
controllable quantity. For the Taylor state $\Theta = 
\lambda a/2 $.

The details of any relaxed state are determined by two 
physically meaningful
parameters, the toroidal flux and the volts-seconds of the 
discharge.  The
toroidal flux as defined  earlier serves to determine the 
field amplitude
and the volts-seconds describes the helicity of the relaxed 
state through
the relation: ${\rm {volts-sec =  helicity}}/{\rm {flux}}^2 
$.
We therefore calculate the helicity integral (global 
helicity) from our solution for the $m=0,~k=0$ state and get 
\begin{eqnarray}
K &=& 4 \pi^2 \alpha_1^2 \frac{R}{a} {(\lambda a)}^3 
\left [ J_0^2(\lambda a) + J_1^2(\lambda a) - 
\frac{2}{\lambda a} J_0(\lambda a) J_1(\lambda a)
\right ]  \nonumber \\
&+& 2 Re \left [ 
\frac{\alpha_2^2}{\alpha_1^2}\left[J_0^2(\lambda\omega 
a)+J_1^2(\lambda\omega a) 
- \frac{2}{\lambda \omega a} J_0(\lambda \omega a) 
J_1(\lambda \omega a)
\right] \right ] \nonumber \\
&+& 2 Re \left [ \frac{\alpha_2}{\alpha_1}\frac{2}{\lambda 
a (1-\omega^2)}
\left [ J_0(\lambda a) J_1(\lambda \omega a)-\omega 
J_1(\lambda a)
J_0(\lambda \omega a) \right] \right ] \\
&+& 2 \frac{{\mid \alpha_2 
\mid}^2}{\alpha_1^2}\frac{1}{\lambda a (1-\omega)}
\left [ \omega J_0(\lambda \omega^2 a) J_1(\lambda \omega 
a)-J_1(\lambda \omega^2 a)
J_0(\lambda \omega a) \right] 
\nonumber \end{eqnarray}

We then calculate the volts-seconds for the cylindrically 
symmetric state
 using $\lambda a = 3.11 $. 
 For the minimum value of $\lambda a = 3.11 $, the
critical value of  volts-seconds = 12.8 $ {R}/{a} $.
For values of volts-seconds less than this critical value,  
a lower value of $ \lambda a$ is obtained from solving the 
equation for $K/\Phi_z^2 $ so that
the cylindically symmetric state is the relaxed state for 
minimum
energy dissipation.  For values of volts-seconds greater 
than the 
critical value the system relaxes to the helically 
distorted state 
with $\lambda a = 3.11 $ which is 
obtained as a mixture of the $m= 0, k= 0 $ and the $m= 1,k 
\neq 0 $ states
as in the case of Taylor's theory.

The profiles for the current and magnetic field  
are shown in Figs. 1 and 2 for the $m=0, k=0$ state with 
the
value of $\lambda a < 3.11$ (i.e., volts-sec $< 12.8 R/a$).
At a value of $\lambda a $ 
greater than 2.95, the magnetic field profile $B_z $ vs $r 
$ shows a 
reversal near the  edge ( Fig.2 ).  Also, $j_z $ and 
$j_\theta $ go to zero at
the wall because of the boundary conditions we have chosen.
The values of both $F $ and $\Theta $ at the boundary $r = 
a $ are
evaluated and $F $ is plotted against pinch ratio $\Theta $ 
( Fig.3 ).  
It is observed
that $F $ reverses at a value of $\Theta = 2.4 $, ($\lambda 
a = 2.95 $)
whereas for the Taylor
state the reversal is achieved at $\Theta = 1.2 $.  
However, this field reversed state supports pressure gradient in
constrast to the Taylor state.
The q-profile, where $q = rB_z/R B_\theta $ is shown
in Fig. 4 for $\lambda a = 3.0 $.

The pressure profile can be obtained from the relation 
${\bf j} \times {\bf B }= \nabla p $. For the $m=0,k=0$ 
state, the only nonvanishing component of the pressure
gradient exists in the radial direction.
The pressure profile is shown in Fig. 5 for the $m=k=0 $ 
state with
$\lambda a =  3.0 $ which is the minimum energy 
dissipation, 
field reversed state.

To conclude, the principle of minimum dissipation is 
utilized together with the constraints
of constant magnetic helicity to determine the relaxed states of a 
magnetoplasma
not driven externally.  The
variational principle leads to a remarkable Euler-Lagrange
equation, and it is shown that this equation involving higher order curl
operator can be solved in terms of
of an analytical continuation of  Chandrasekhar-Kendall 
functions   in the complex domain with appropriate boundary conditions.  
 This relaxed state obtained from single fluid MHD supports
pressure gradient.  A coupling between magnetic field and flow is not
an essential criterion for having a non-zero pressure gradient.
Further, it is  shown that a non force-free state with field reversal 
properties can exist.

One of the authors (BD) wishes to
acknowledge the kind hospitality of ICTP, Trieste, Italy, 
where a
part of the work has been carried out during his visit as an 
Associate.  The author gratefully acknowledges many illuminating
and inspiring discussions with Predhiman Kaw, David C. Montgomery,
Swadesh Mahajan and Zensho Yoshida.

\newpage
\begin{center}
{\bf {Figure Captions}} \end{center}
\vglue 1 cm
Fig.1~~ $j_z(r)$ vs $r$ for the axisymmetric state 
$m=0,~k=0$, and
$\lambda a=3.0$. The current vanishes at the edge because 
of the 
boundary condition ${\bf j}\times{\bf n}=0$.
\vglue 1 cm
Fig.2~~  Magnetic field profile for the axisymmetric state 
with 
$\lambda a=3.0$, showing field reversal near the edge.
\vglue 1 cm
Fig.3~~ The field reversal parameter $F$ against the pinch
parameter $\Theta$, the field reversal occuring at $\Theta 
=2.4$.
 The dotted curve represents the plot for the minimum 
energy 
 state of Taylor.
\vglue 1 cm
Fig.4~~ The q-profile for the axisymmetric state.
\vglue 1 cm
Fig. 5~~ The pressure profile $p$ vs $r$ for $\lambda 
a=3.0$.

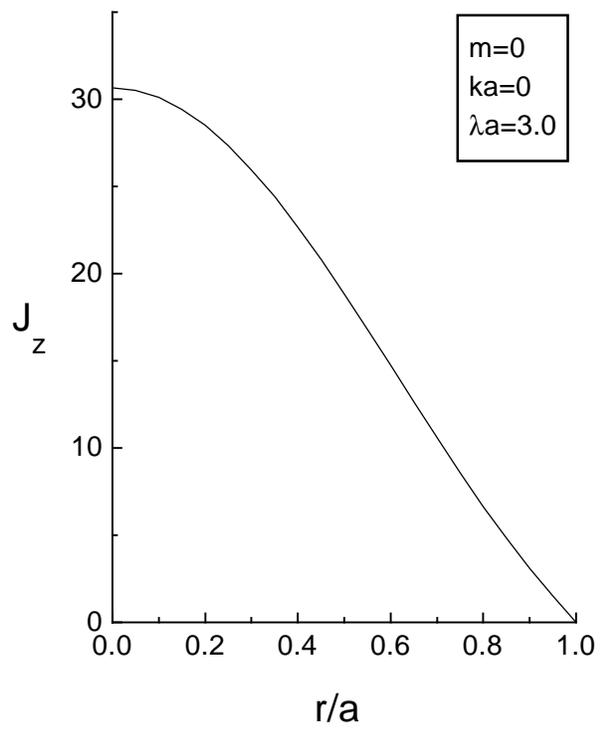
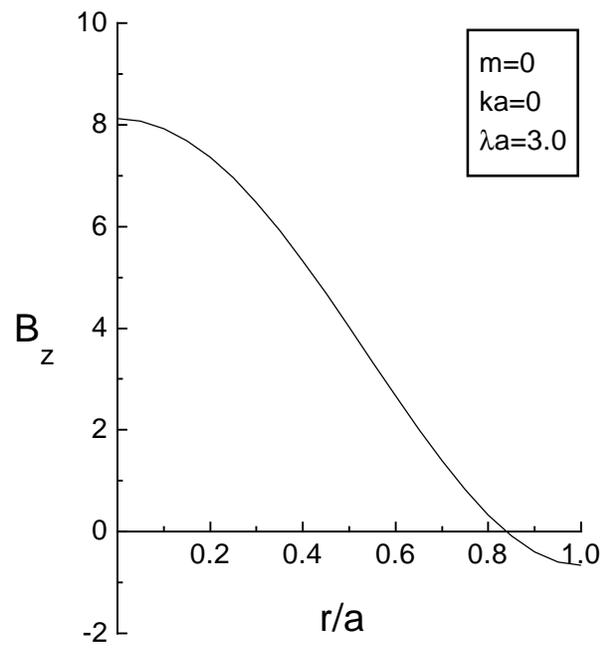
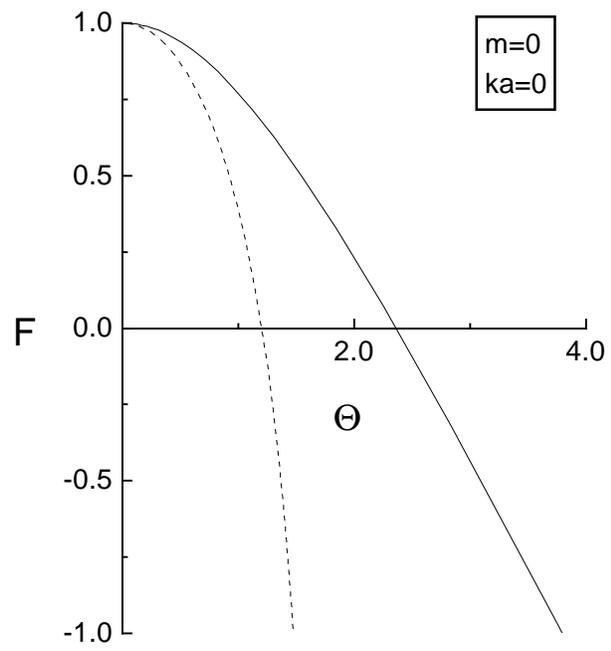
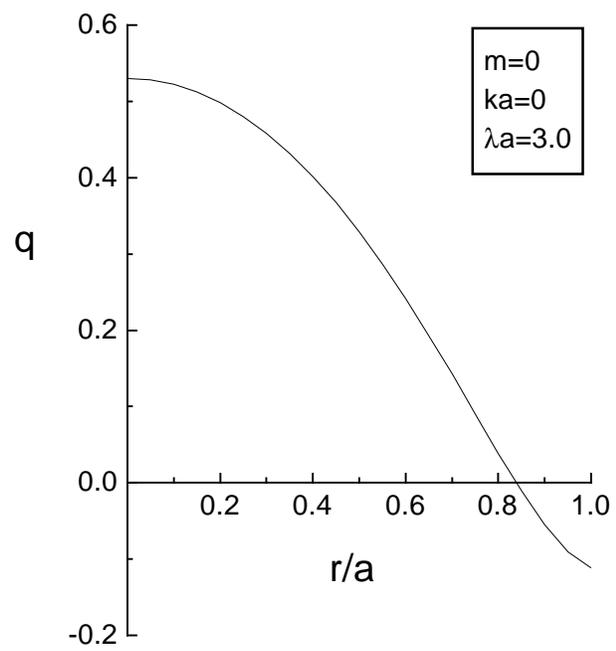
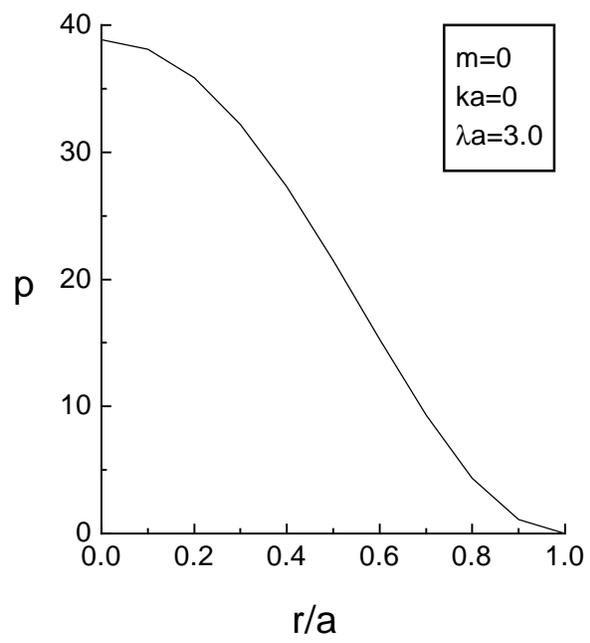
\end{document}